\input{psfig.sty}

\documentclass{XrU2005}

\usepackage{setspace}

\title{X-ray flare modeling in the single giant HR 9024}
\author[1]{Paola Testa}
\affil{MIT Kavli Institute for Astrophyisics and Space Research, 02139 Cambridge MA, USA}
\author[2]{David Garcia-Alvarez}
\affil{Harvard-Smithsonian Center for Astrophyisics, 02138 Cambridge MA, USA}
\author[3]{Fabio Reale}
\affil{Dip. Scienze Fisiche ed Astronomiche, Universit\`a di Palermo, Palermo, Italy}
\author[1]{David Huenemoerder}

\begin{document}

\keywords{High resolution X-ray spectroscopy, hydrodynamic loop models}

\maketitle

\begin{abstract}
\vspace{-0.5cm}
We analyze a {\em Chandra} HETGS observation of the single G-type giant
HR 9024. The high flux allows us to examine spectral line and continuum diagnostics
at high temporal resolution, to derive plasma parameters (thermal distribution, 
abundances, temperature, ...). A time-dependent 1D hydrodynamic loop model with 
semi-length 10$^{12}$cm ($\sim R_{\star}$), and impulsive footpoint heating triggering 
the flare, satisfactorily reproduces the observed evolution of temperature and 
emission measure, derived from the analysis of the strong continuum emission.  
The observed characteristics of the flare appear to be common features in very large 
flares in active stars (also pre-main sequence stars), possibly indicating some 
fundamental physics for these very dynamic and extreme phenomena in stellar coronae.
\end{abstract}

\vspace{-0.5cm}
\section{HETGS observations and analysis}

\vspace{-0.5cm}
\begin{figure}[!hb]
\centerline{\psfig{figure=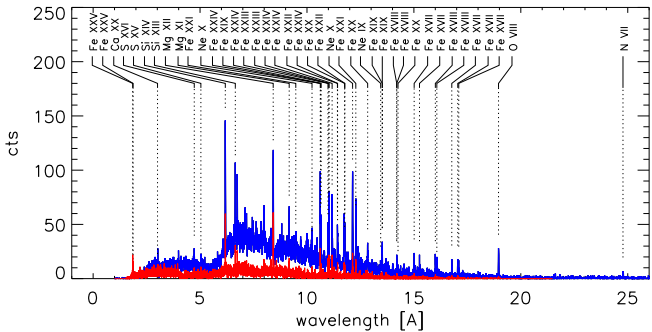,width=9.5cm}}
\vspace{-0.5cm}
\caption{{\em Chandra} HEG (red) and MEG (blue) spectra obtained
in a $\sim 96$~ks observation of the single giant HR~9024.\label{fig:spectrum}}
\end{figure}

\begin{figure}
\centerline{\hspace{-0.5cm}\psfig{figure=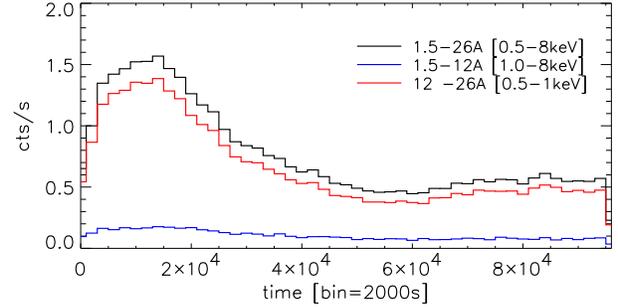,width=9cm}}
\vspace{-0.5cm}
\caption{Lightcurve obtained as the sum of total counts of HEG and
MEG dispersed spectra. Lightcurves in hard (red) and soft (blue) spectral
bands are shown. \label{fig:lightcurve}}
\end{figure}

Figure~\ref{fig:spectrum} and \ref{fig:lightcurve} show the {\em Chandra} High
Energy (HEG) and Medium Energy (MEG) Gratings spectra, and the lightcurve for 
a $\sim 96$~ks observation of the single giant HR~9024.
The lightcurves for a hard and a soft spectral band, shown in Fig.~\ref{fig:lightcurve},
indicate an harder spectrum during the two peaks of the emission ($\sim 15$~ks 
and $\sim 80$~ks from the start of the observation), typical of stellar flares.
Table~\ref{tab:table} summarizes the characteristics of source and the parameters
of the {\em Chandra} observation.

\vspace{-0.5cm}
\begin{table}[!hb]
  \begin{center}
    \caption{Stellar parameters and HETGS observation.}
    \renewcommand{\arraystretch}{1.2}
 \begin{footnotesize}
    \begin{tabular}[h]{cccccccc}
      \hline
 Spec. &  d   & $R$	&  $M$  &  $\log L_{\rm bol}$  &  $P_{\rm rot}$ & $\log L_{\rm X}$ & $t_{\rm exp}$ \\
     type & [pc] & [$R_{\odot}$] & [$M_{\odot}$]  &  [erg/s]   &  [days] & [erg/s] & [ks]\\
      \hline
  G1 {\sc iii} & 135 & 13.6 & 2.9 & 35.4  & 23.25 & 31.8 & 95.7\\
      \hline \\
      \end{tabular}
    \label{tab:table}
 \end{footnotesize}
  \end{center}
\end{table}

\vspace{-1.2cm}
\paragraph{Spectral Analysis:}

The high resolution spectra provide several plasma diagnostics, from the analysis 
of both continuum and emission lines, and from the lightcurves in different spectral 
bands or in single lines.
The evolution of temperature and emission measure (EM) during the flare allows to 
construct a model of the flaring structure(s) (Reale et al.\ 1997). 
\vspace{-0.5cm}
\begin{itemize}
 \item T is derived from the fit to the continuum emission, selecting spectral regions 
 	"line-free" (on the basis of predictions of atomic databases such as APED [Smith 
	et al.\ 2001], CHIANTI [Dere et al.\ 1997]); the fit also provides an estimate 
	for EM from the normalization parameter.
 \item The emission measure distribution (DEM) is derived through a Markov-Chain 
 	Monte-Carlo analysis using the Metropolis algorithm (MCMC[M]; Kashyap \& Drake 1998) 
	on a set of line flux ratios (O lines are the coolest Ar the hottest, i.e.\
	logT[K]~6.2-7.8). Coronal abundances are evaluated on the basis of the derived DEM; 
	the abundance is a scaling factor in the line flux equation to match the measured flux. 
\end{itemize}

\vspace{-0.7cm}
\section{Results}
\begin{figure}[!h]
\centerline{\psfig{figure=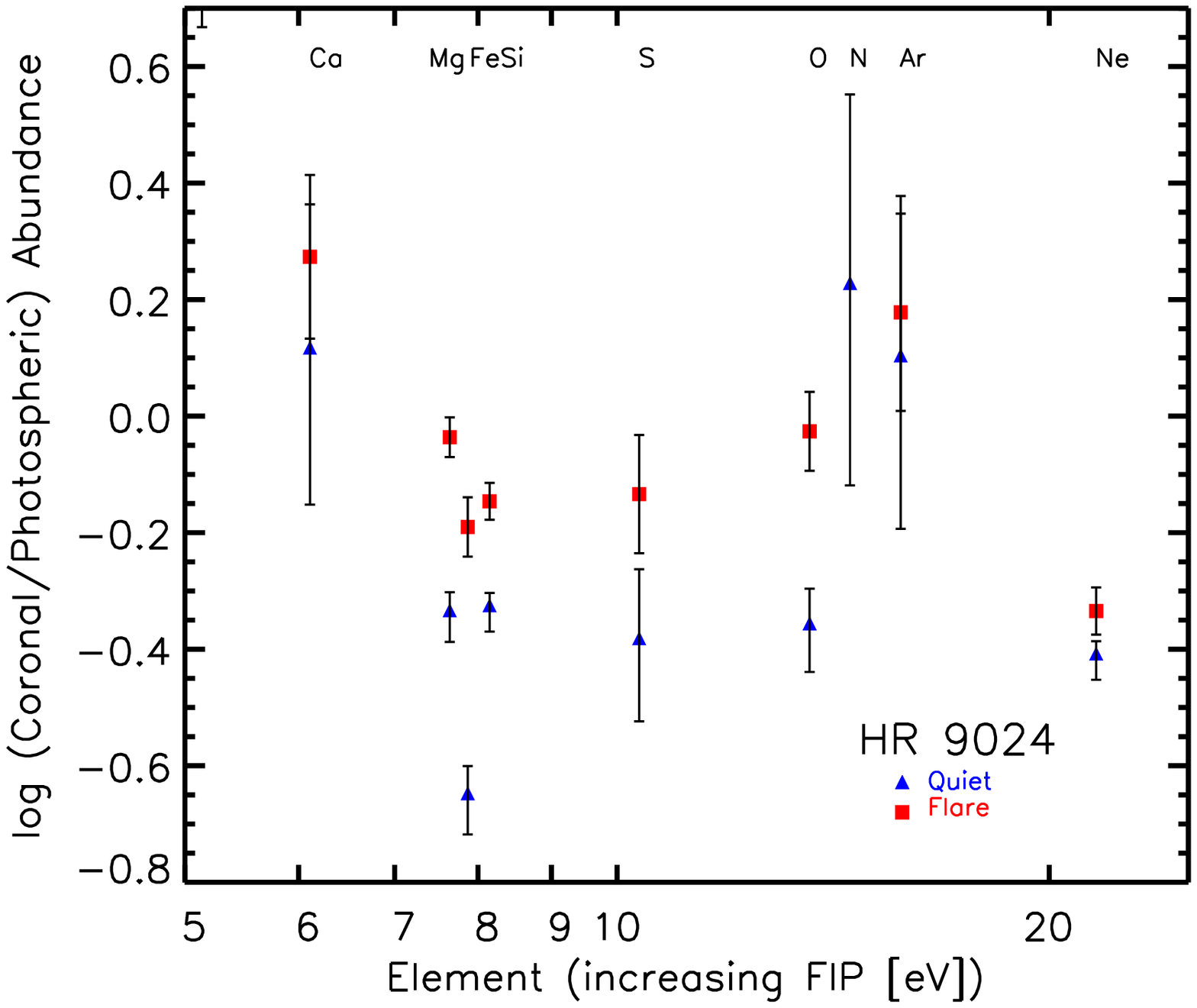,width=4.5cm}
\psfig{figure=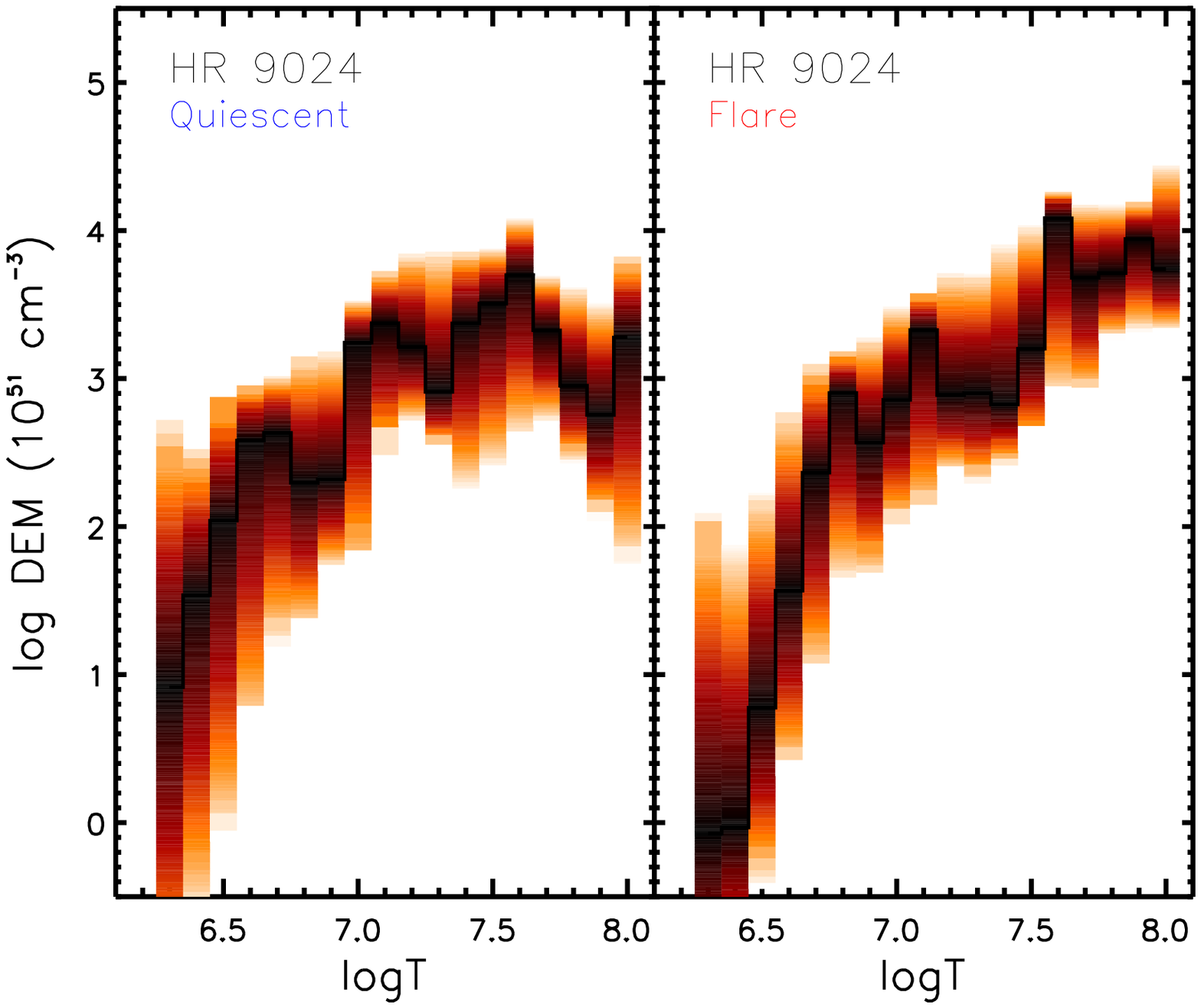,width=4.5cm,height=3.6cm}}
\vspace{-0.3cm}
\caption{Abundances ({\em left}) and DEM ({\em right})
derived for the flare and the quiescent phase.\label{fig:dem_abund}}
\end{figure}

\vspace{-0.5cm}
\paragraph{Abundances and DEM:} 
Fig.~\ref{fig:dem_abund} show the abundances ({\em left}) and DEM ({\em right})
derived for the flare peak, and the quiescent emission.
\vspace{-0.5cm}
\begin{itemize}
 \item[-]  abundances variations between flaring and 'quiescent' phases \vspace{-0.25cm}
 \item[-] very hot corona, also outside the flaring phase, as found also from an 
 	{\em XMM-Newton} observation showing no obvious flare (Gondoin 2003)
\end{itemize}

\vspace{-0.5cm}

\vspace{-0.7cm}
\paragraph{Hydrodynamic Modeling:} We aim at reproducing the observed evolution of 
temperature, T, and emission measure, EM.  
The 1D hydrodynamic model, solves time-dependent 
plasma equations with detailed energy balance; a time-dependent heating function
defines the energy release triggering the flare. The coronal plasma is confined in 
a closed loop structure: plasma motion and energy transport occur only along magnetic 
field lines.
Parameters: (1) loop semi-length $L=10^{12}$~cm (a first estimate of L is obtained 
from the observed decay time); (2) footpoint heating; (3)  initial atmosphere: 
hydrostatic, $T=2 \times 10^7$~K; however, the initial conditions do not affect the 
evolution of the plasma after a very short time.
 
\begin{figure}
\centerline{\psfig{figure=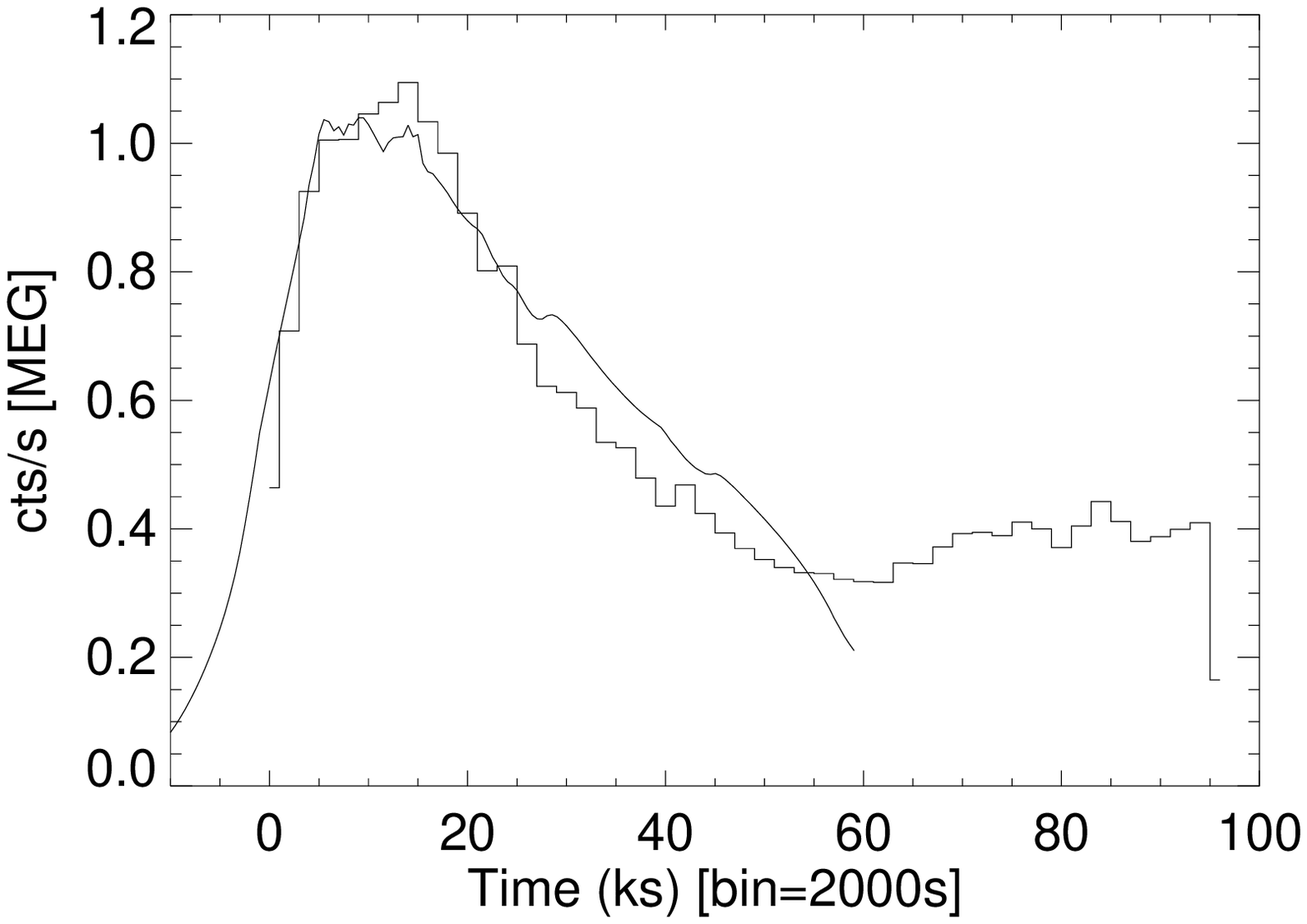,width=5cm}\hspace{-0.4cm}
\psfig{figure=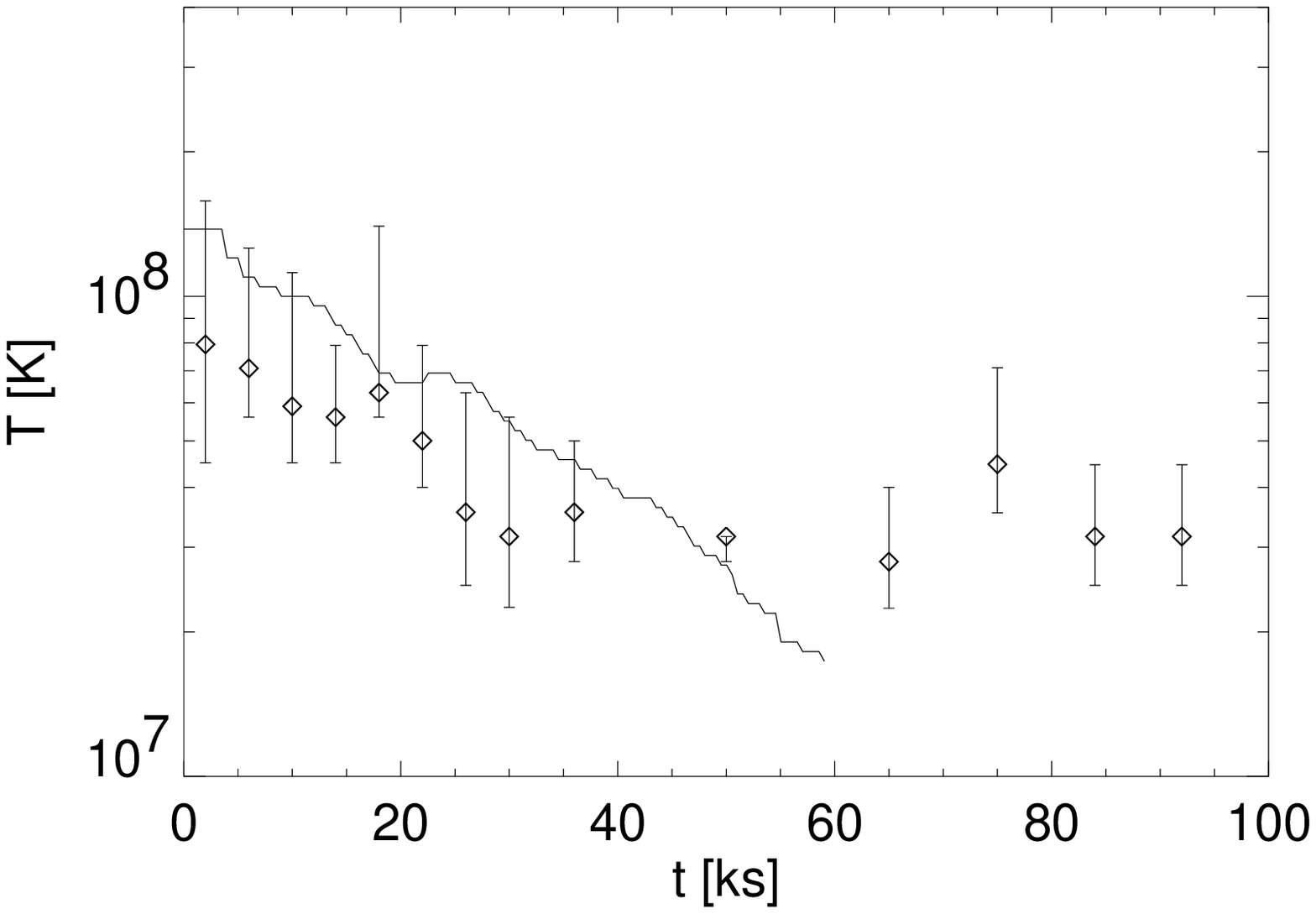,width=5cm}}
\centerline{\psfig{figure=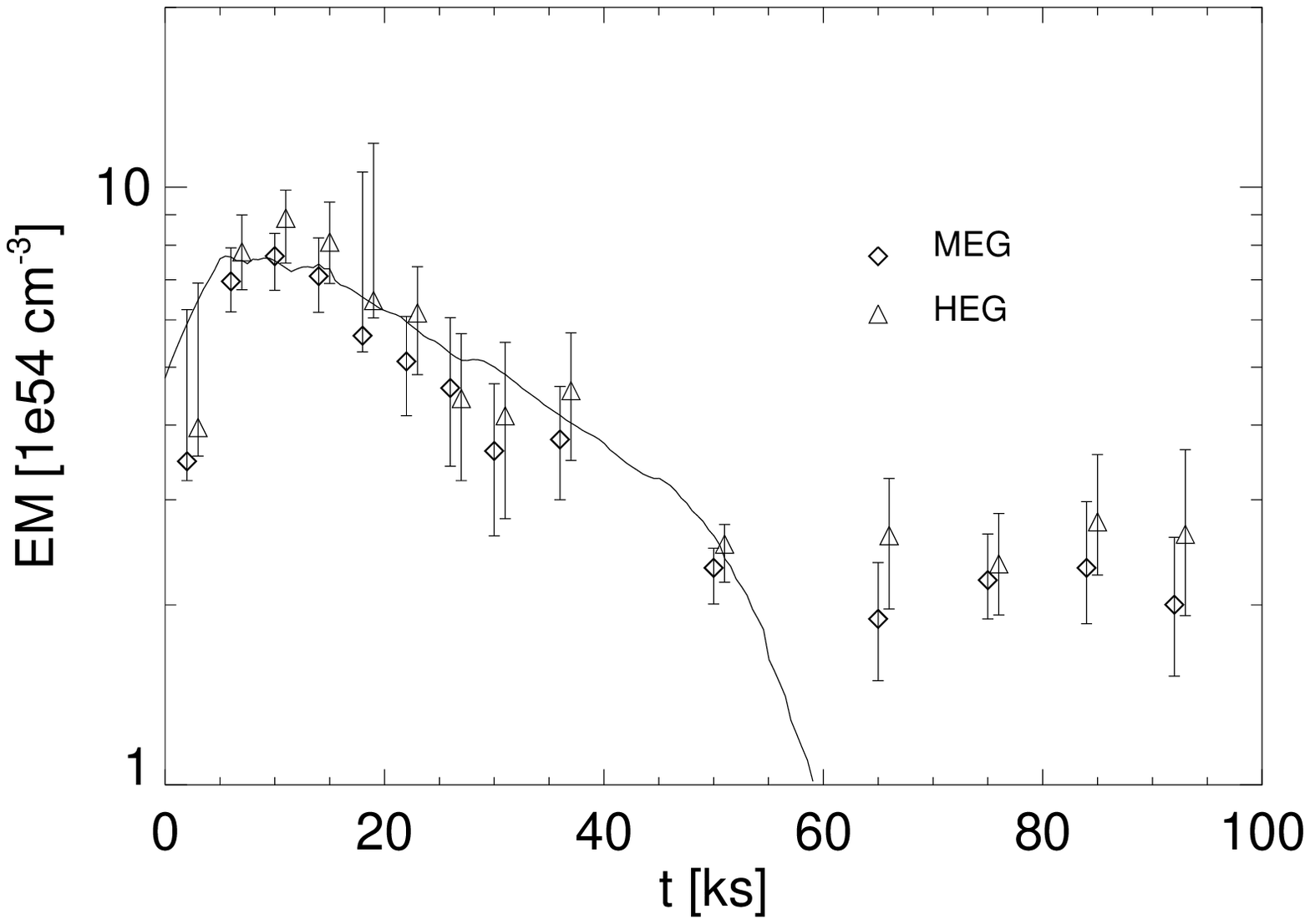,width=5cm}\hspace{-0.4cm}
\psfig{figure=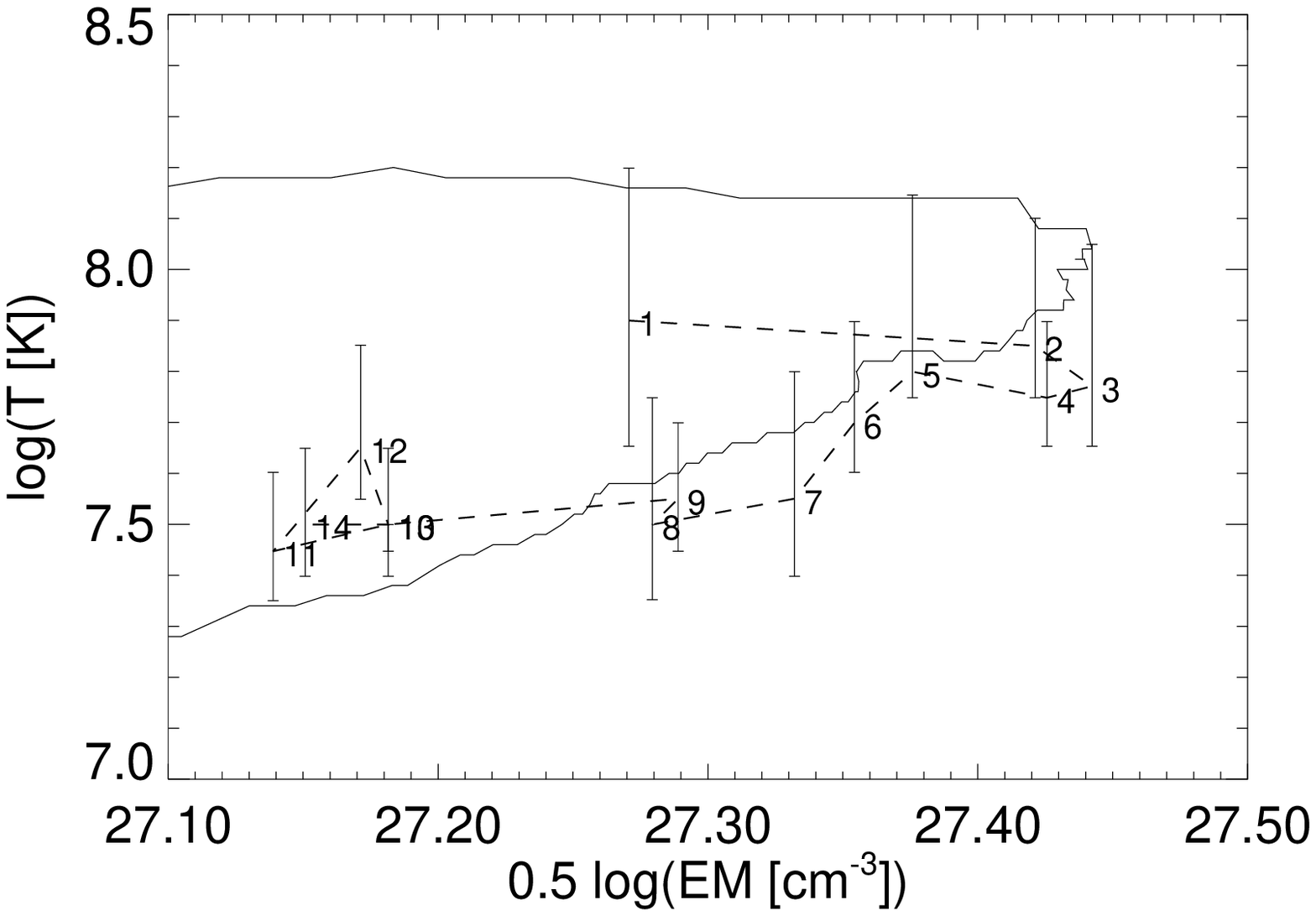,width=5cm}}
\vspace{-0.5cm}
\caption{Comparison of observed lightcurve (top left) T (top right),
	EM (bottom left), and T-n (bottom right) evolution, with the 
	corresponding quantities synthesized from the hydrodynamic
	model (solid lines).\label{fig:modelfit}}
\end{figure}

Fig.~\ref{fig:modelfit} show the comparison of observed T and EM evolution, and 
X-ray lightcurve, with the corresponding quantities synthesized from the 
hydrodynamic model.

The observed evolution is reproduced reasonably well by a model characterized by:
\vspace{-0.5cm}
\begin{itemize}
 \item[-]  loop semi-length $L=10^{12}$~cm ($\sim R_{\star}$);\vspace{-0.25cm}
 \item[-]  impulsive (20~ks, shifted by 15~ks preceding the beginning of observation) 
 	footpoint heating triggering the flare; no sustained heating (i.e.\ pure cooling);\vspace{-0.25cm}
 \item[-]  volumetric heating $\sim 10$~erg/cm$^3$/s, heating rate $\sim 8 \times 10^{32}$~erg/s;\vspace{-0.25cm}
 \item[-]  from the normalization of the model lightcurve we derive an estimate of loop 
 	aspect ratio $\alpha=r/L \sim 0.023$, i.e.\ the loop cross-section has radius 
	$r \sim 2.3 \times 10^{10}$~cm.
\end{itemize}

\vspace{-0.8cm}
\section{Conclusions}
\vspace{-0.5cm}
The evolution of this very hot spectrum is reproduced by an hydrodynamic loop 
model with $L \sim R_{\star}$. This loop model has roughly the same parameters
of models satisfyingly reproducing other large flares: e.g.\ flares in pre-main 
sequence stars (Favata et al.\ 2005). Large flares observed in very active stars 
seem to have very similar characteristics, possibly with important implications 
for the physics of these phenomena.
Large flares as the one observed for HR~9024 are very unusual in single evolved
stars, while being more common in active binary system.

Future work:
(1) finer spectroscopic analysis, and hydrodynamic modeling; detailed comparison of 
 	observed spectra with synthetic spectra derived from hydrodynamic model. The high 
 	resolution spectroscopy together with the high signal for this observation provides 
	a large amount of constraints to the model. 
(2) explore possible evidence of Non Equilibrium Ionization effects.
(3) determine robust constraints on abundance variations during flare
(4) analysis of Fe fluorescent emission: we can obtain constraints on the geometry 
 	of the emitting plasma, in particular on the height of the illuminating source, 
	i.e.\ the loop sizes, obtaining a cross-check to the results of the hydrodynamic 
	modeling.   This observation provides the first clear evidence of fluorescence 
	in post-PMS stars other than the Sun (i.e.\ fluorescence from photosphere; while 
	in PMS stars there is evidence that the fluorescence emission is coming from the 
	accretion disks).

\vspace{-0.4cm}
\small
\begin{spacing}{0.85}

\end{spacing}

\end{document}